# Model Based Design Environment for Launcher Upper Stage GNC Development

Hans Strauch[1], Klaus Luig[1], Samir Bennani[2]

[1]Airbus DS GmbH
Airbus Allee , 28361 Bremen
Germany
Email:hans.strauch@airbus.com. Klaus.Luig@airbus.com

[2]ESA/ESTEC
Keplerlaan 1, 2201 AZ Noordwijk
The Netherlands
Email:samir.bennani@esa.int

**INTRODUCTION**

The cost associated with developing flight control forms a significant part of the overall development cost. The increased demand for greater functionality and for extending the domain of applicability of future launchers leads to higher complexity of the GNC algorithms. Furthermore, there is also a need for a responsive design methodology, which can quickly adapt to changes in the requirements during the development process. These demands are stretching the current, largely manual, development process, which is fragmented in the different disciplines and activities concerning modelling, algorithm design, software coding, implementation on the avionics platform and the associated testing at the different stages.

The model-based-design (MBD) philosophy presents an attractive solution in addressing the multi-disciplinary nature of the flight algorithm design task. In terms of design process, this materializes in a tight coupling between the modelling, the design and the analysis activities. An integrated design framework for the GNC development for future launcher upper stages has been realized with the 'Upper Stage Attitude Development Framework' (USACDF) research effort. USACDF has been initiated by the ESA Launcher Directorate within its Future Launcher Preparatory Program (FLPP). FLPP aims at preparing the Next Generation Launcher (NGL) by supporting activities, which increase the maturity level of launcher technologies and acquire new ones. It specifically implements a system-driven approach.

The framework seamlessly covers, in a dynamic fashion, the entire design process from requirement capturing activities until verification and validation of the auto-coded GNC and the related mission and vehicle management (MVM) application software, which runs on flight representative processors and hardware (LEON II used for demonstration).

The basis of the design suite relies on the commercially available MathWorks' tool-chain accompanied by the dSPACE system (see [1], [2]). However, it is not limited to the algorithm design, but also heavily using those elements, supporting model based design, automatic report generation, requirement tracing between software and specifications and automated verification, including real-time processor performance profiling. Mathworks' GNC and SW tool-chain has been augmented to physical modelling approaches, namely Modelica and EcosimPro (beside Mathworks' own SimMechanics). The particular strength of the physical modelling is the injection of failure cases directly at the physical level. This allows for designing and testing advanced failure recovery elements of the flight software.

The algorithms, developed in the USACDF framework, can be run in a TASTE/RASTA/LEON environment (see [3]) via autocoding. The plant dynamics is realized on a dSPACE system connecting to RASTA via Ethernet communication. The transfer from a PC based functional engineering simulator (FES) to a LEON/dSPACE closed loop configuration can be achieved quickly and nearly automatically.

The paper is organized as follows: The next section will place the USACDF approach into an overall context by discussing contributions found in the literature. Next, the limitations and constrains of the current way to develop flight control for space application are discussed. The following sections will then highlight the USACDF design suite features and explain how they answer to the needs and how they extent the domain of applicability. The last section discusses the benchmark problem, which was used to test the framework and to carry out algorithm design for some particular uppers attitude control problems.

**RELATED WORK**

The need to enable effective collaboration of people across multiple disciplines and organize them during the integration and testing phase has been identified early on. Ref [4] discusses this topic and highlights the need for increased parallelization especially for the testing phase. Ref [5] drives the approach for this phase even further by discussing automated test generation combining UML and the MathWorks toolchain. Ref [6] concentrates on the model-based design strategies for real-time hardware-in-the-loop simulations for flight code validation. While above papers deal with the way of how to efficiently organize different disciplines and groups of engineers and how to improve the process of interaction, [1] discusses the improvement potential by using graphical tools for function specification. This paper shows that MathWorks was an early pioneer in promoting the model-based paradigm and transformed its toolchain accordingly. Ref [7] and [8] show that the concept has been actually adopted in the recent development of the Orion GNC development and already presents "lessons learned".

After the wide adoption of the MBD principle, activities are under way to standardize the ideas and fit them into the known standards like DO-178 (see e.g. [9]). While MBD is a general approach, space approved flight software has its town particularities, which is discussed in [10]. The references above have been mainly concerned with the processes and the tools. However, MBD can also influence and improve the GNC algorithms themselves. Ref. [11] shows how autonomy (termed "self-management") and online reconfiguration capable system can be better developed within the MBD concept. The study presented in this paper also went into that direction by illustrating the improvement potential of MBD applied to a particular difficult flight phase of an upper stage ("long coasting phase"). The next section gives a detailed description of the needs and how they are answered by the MBD principle.

**CURRENT GAPS, LIMITANTIONS AND NEEDS**

As already indicated in the discussion of the related literature, it is a common perception that the current way of development is not sufficiently adapted to the challenge. In the following we will provide an overview of the shortcomings of the current way of working and highlight the needs. Fig 1 illustrates the current situation and contrasts it with the model-based design (MBD) approach. The involved disciplines, like structure, propulsion, software, etc., are usually working within separated analysis environments and with a wide collection of tools, with only limited interfacing options ("throw-it-over-the-wall process" [4]). Creating a combined simulation model, which is adapted to the needs of the GNC algorithm development, is tedious work and the degree of model complexities varies over the disciplines.

What is needed are models suited for closed loop simulation. The option for closed loop simulation is important, because running the models in the closed loop assures that the models will be excited in the right input domain (signal pattern, frequency, amplitude, etc.). For example, the thrust from a cold gas propulsion system depends on the pressure in the tank, which depends on the attitude orientation (thermal impact). On the other hand, the attitude evolution is determined by the controller, but its performance depends on the variation in the thrust level. Currently, dependencies like that can be evaluated only very late, if at all. In the beginning of the design cycle the models must be lean and quick to execute, but later they must be capable of gradually increasing the model fidelity, as soon as results from hardware subsystem tests become available. This tight connection and the option to upgrade are not only important for speeding-up the design, but also for identifying needed GNC algorithm improvements as early as possible.

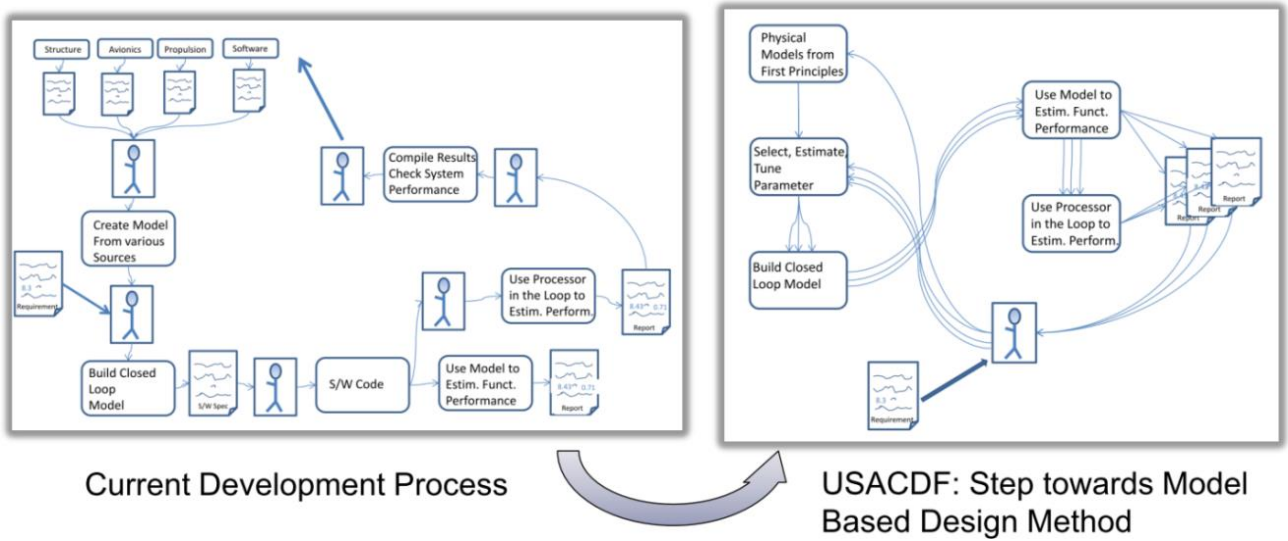

Fig 1. Current fragmented, manual intensive Process (left) and unified, multi-disciplinary, effective MBD Approach

The activity of setting-up a unified, multi-disciplinary modelling environment with a tight link to the validation via subsystem hardware tests, will implicitly move the subsystem engineers closer to the overall system engineering needs and will lead to better adapted models for use at system level.

The GNC development activity plays a pivotal role in this task. The desired GNC performance can only be achieved with the right combination of sensors, actuation system, understanding of the environmental constrains and limitations in software and avionics. Therefore GNC design is inherently multi-disciplinary and, as such, one of the closest subsystem to the overall system engineering. This is illustrated in Fig 2. GNC is placed at a quite central position in the bottom figure, actually establishing the link not only between most of the subsystems but also between development activities: System definition, S/W development and testing.

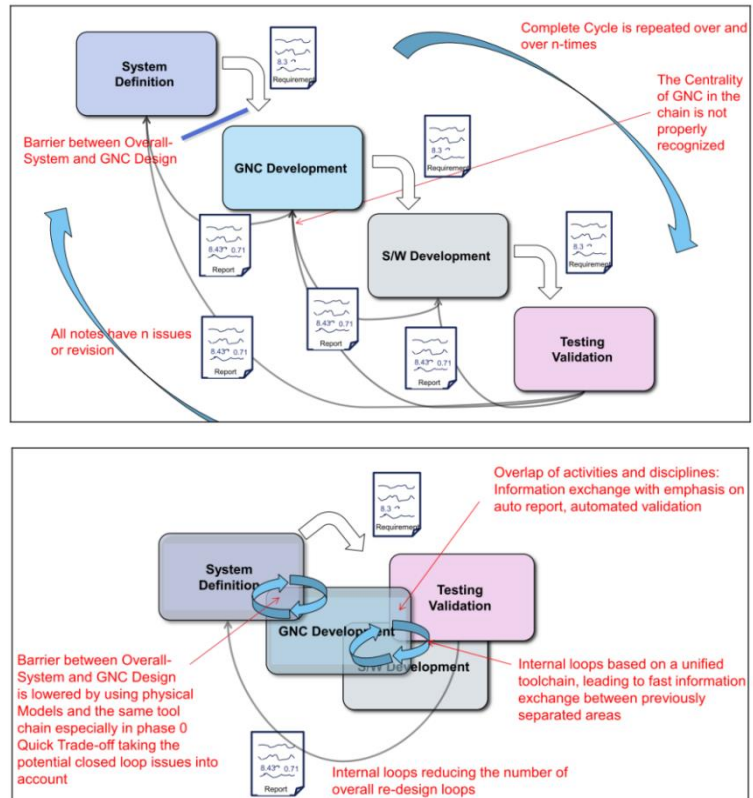

Fig 2. Current sequential, lowly connected design cycle (top). Strongly integrated, overlapping, partly concurrent process in the MBD approach centered on GNC (bottom).

In the following we will list, on a more detailed level, the gaps and constraints in the current methodology:
1. **System Oriented GNC Design** Due to the separation and sequentiality of the classical design phases, the link and flow-down of the requirements on system level to GNC subsystem requirements is a laborious and failure prone process. What is missing is the option for a quick check, early in the development state, on the preliminary GNC requirements set, allowing for fast iteration with the system specification. As the GNC performance is heavily influenced by the choice on sensors, actuators and environmental constrains, a quick closed loop simulation capability can support the system design. This centrality of GNC in upper stages is naturally accounted for in a MBD approach.
2. **Modelling Domain** The current way we model, leads to limitations on what we can achieve. The external environment and the interdependency of structure, actuator, propellant location, thermal condition and sensor physics is a kind of closed loop in itself. It must be adequately reflected in the composite model onto which the GNC algorithm is build. Especially for the attitude dynamics of upper stages the thermal condition and the evolution of thrust, used for actuation, is tightly connected via the physics of thermal, fluid and orbital dynamics. This is not reflected in current simulators. The effects are phased-in iteratively via open loop profiles for thermal and fluid. The USACDF development suits addressed this issue. A further example is the separation of payloads. Usually the separation process is modeled in the current closed loop simulators as an equivalent delta v, augmented with some uncertainties in magnitude and direction. These parameters are evaluated in dedicated studies. The detailed physics is simulated separately and only phased-in partially into the end-to-end simulator. With a physical modelling approach the details of the separation process, including the electrical command chain, can be captured directly and the consequences of failures in the way of interaction with GNC commands can be studied in a representative way. USACDF has addressed this problem via the Modelica simulation of the full separation chain.

3. **Operational Robustness** The degree of operational robustness is small in current launchers. Although redundant actuators are not common on upper stages, there are still options for mitigating actions, provided that MVM is flexible enough to adapt in real-time. In order to achieve this, three new elements are needed: An appropriate way of specifying adaptive mechanism in mission definition languages, the corresponding simulation capability of inducing failures into the closed loop execution and verification methods for building a sufficient confidence in the algorithm. The first need is answered by the programming language of higher abstraction like state machine, temporal logic, etc. The second need is covered by physical modeling and the third need is covered by automated and formal methods, like Polyspace analysis. USACDF has addressed this issue with StateFlow as a mean of defining the nominal, as well as failure recovery, payload release sequence
4. **Avionics Resources** Modern control techniques have the promise of extending the domain of adaptations, responsiveness and robustness of upper stage control. However, they will often increase the demand on the hardware resources. Currently the demand on avionics resources establishes itself only lately in development phase. A capability of achieving processor-in-the- loop (PIL) early is missing. In USACDF a nearly automated process for moving, via autocode, to PIL on flight representative hardware has been achieved.
5. **Failure Handling** Currently there is only a limited amount of failure recovery in upper stages. Although the hardware resources for real redundancies are limited in launchers, there are still options to cope with hardware degradations providing one can adapt flexible, automated and in real-time the MVM and retune the control algorithm (e.g. jeopardizing performance for recovering a certain loss in gain and phase margin due to unexpected, but identifiable uncertainties). A complete physical modeling of the full command chain, which can be used in closed loop, is often missing. Each failure case currently demands an extra, dedicated set of differential equation, which must be set-up manually. A unified modelling approach, covering nominal and off-nominal, is needed to build-up confidence into GNC based failure response features. USACDF provided some improvements by employing physical modelling connected with partly automated V&V features.
6. **Autonomy** Improved operational robustness and failure handling capabilities are part of an overall increase of autonomy. The current models, simulation and applied control methods are tuned only to sequential and deterministic execution of the mission sequence. New and improved elements in terms of physical models, integrated simulation tools and modern, adaptive control methods already exists and they are employed in other application areas successfully [2]. The presented study results of USACDF are a step in introducing these techniques into the launcher development environment.
7. **Requirement Process** Establishing requirements from the system down to the various subsystems is a laborious process, involving many engineers, many hand written documents and many loops to weed-out failures in the interfaces. Working in an integrated way, based on the principals of the physics involved, this process can be sped-up, if tools are available, which supports the linking of the requirements with the software and provide automatic reports. Such an environment also encourages a tree like structure of the requirements, as opposed to separated and flat sets of requirements currently given for each subsystem. In USACDF a benchmark launcher test case has been set-up, based on such an approach using various Mathworks toolboxes for requirement handling, linking and reporting.
8. **Domain of Applicability** New systems often demand a wider or different domain in which the vehicle must operate. In the current approach, extending the domain demands lots of additional work, incorporating various groups on engineers. MBD can prove this situation by having already the mechanism of modelling and interfacing, in a general and unified way, in place. Once the physics of the particular, new phenomenon has been set-up in the relevant model, the already established process and tool chain is automatically applied. In USACDF this was tested by working on a control technique for the long coasting phase, which demands a control algorithm taking into account simultaneously the effects of minimum impulse bit, large amount of propellant and certain phenomenon of attitude dynamics like nutation and precession. The current development approach was not able to cover this adequately.
9. **Closed Loop, Multi-Physics Simulation** It has been already sufficiently emphasized that closing the loop is necessary. The current situation often does not allow this. For example the internal sloshing phenomenon is often simulated by a constant, body fixed torque, simply because of the lack of the capability to use CFD simulation in closed loop with the control algorithm. Another example is the following: First a simple rigid body simulation of the mission is performed, the attitude profile is provided to the propulsion department for analyzing the propellant motion along the tank wall, this profile is provided to the tank development group for establishing the temperature and pressure profile. The pressure profile defines the available thruster force as function of time. This profile must be fed-back to the GNC simulation and will change attitude evolution. The cycle must be repeated, but without guarantee of converging. In an MBD tool chain the sloshing CFD simulation is connected with thermal ESATAN simulation and the system is in closed loop with the flight software. It is automated, repeatable and rigorous and it can be employed in the early stages.

# OVERVIEW OF MODEL BASED DESIGN IN USACDF

The model-based concept encompasses the whole activity from first feasibility analysis to final testing and verification. As such it is a wide area and the study had to concentrate on some aspects which were:
- A seamless approach from Simulink based attitude control algorithm via autocoding to processor-in-the-loop
- Physical modelling for some subsystems which are special for upper stages
- Unified requirement process, testing and documentation.
- System identification and adaptive control

Fig 3 illustrates the main elements. The left hand sided figure shows that a part to the plant dynamics has been realized with Modelica (separation process) and EcosimPro (attitude thruster and main engine). It also shows elements dealing with the analysis. They are worst case case/cross entropy analysis, visualization and automated generation of the payload release sequence.

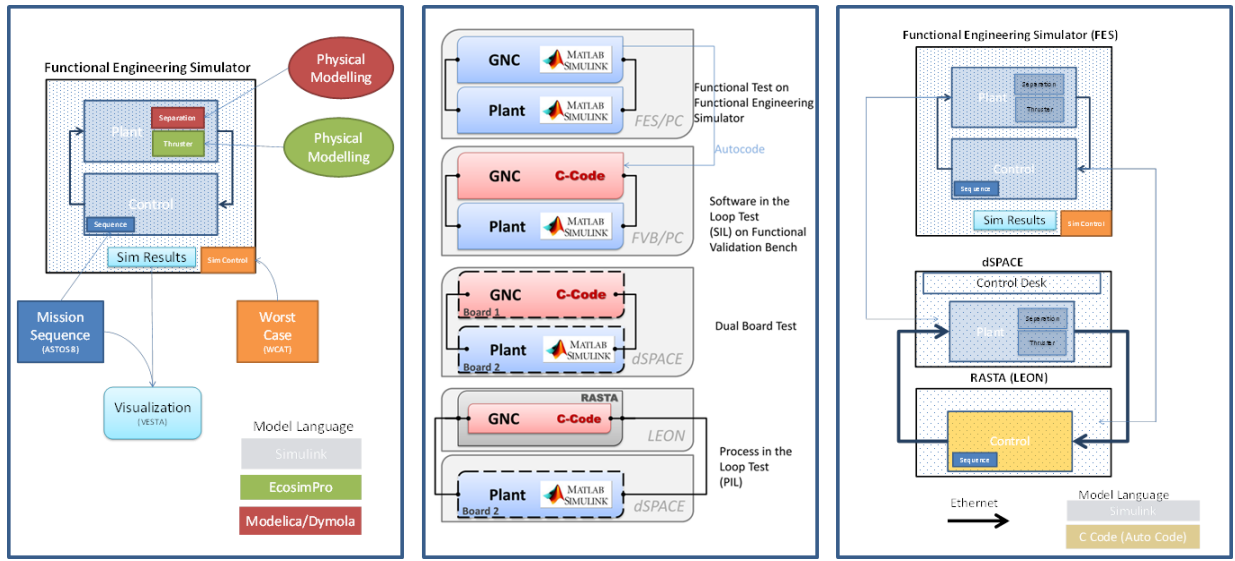

Fig 3. Left: Functional Representation of the Simulator illustrating Modelling Languages and off-line Tools. Middle: Illustration of the Process from deriving PIL. Right: Hardware Configuration with dSPACE and LEON in closed loop via Ethernet.

The figure placed in the middle of fig 3 illustrates the autocode process. The software is first developed in the graphical language of Simulink. The application software is then converted via the embedded coder of MathWorks to C and tested (software-in-the-loop). Very often the exact mechanism of the analog to digital (AD) and the reverse (DA) links are not fully and correctly captured in the Functional Engineering Simulator (FES). Therefore it is important to quickly move to a dual board implementation or even to a flight representative processor. This is shown in the right hand sided figure. Within the USACDF framework the testing on a LEON can be done very early in the algorithm design phase because the process of code generation is nearly automated.

The physical modelling approach has been applied in a hybrid configuration. EcosimPro is firmly rooted in some areas of the propulsion development. The framework succeeded to link the already available models into the Simulink closed loop simulation. The electrical command chain and the physics of the payload separation have been modelled with Modelica. These physical models easily allowed implementing failures (leakage, separation spring uncertainties, delays, etc.).

As discussed in [1] MathWorks has augmented their original set of algorithmic toolboxes with model-based features, namely the toolboxes: Report Generator, Verification and Validation, Design Verifier, Simscape, SimMechanics. In the USACDF framework the features allowing requirement linking between specification documents and software are used. Requirements and software definition reports are generated automatically.

The framework was then applied to several benchmark cases representing the anticipated range of future uppers stage applications. That way the framework was tested while it was build. One application was the adaptive control of a particular flight phase, namely the "long coasting phase". System identification, adaptive and model predictive control was employed within the MBD paradigm.

# ELEMENTS OF THE DESIGN ENVIRONMENT

The processor-in-the-loop (PIL) configuration consists of a dSPACE with DS1006 processor. The 'ControlDesktop' software provides advanced graphical control features to handle the closed loop simulation. An Ethernet connection is used for the link to a LEON2 GR-CPCI-AT679. The application software on the LEON2 is generated via the 'embedded coder' toolbox from MathWorks. (see fig.4)

The critical design task was to accurately map the synchronization features and the delay of the Ethernet based hardware link in the Function Simulation (FES) based on Simulink. It was possible to achieve this to such a degree that closed loop results on FES and LEON2 revealed differences in the order of $10^{-14}$ only.

The design suite also has an Extended Functional Simulator (EFES) which contains those physical model elements which cannot be used in the real-time based configuration discussed above. The separation mechanism has been modeled with Modelica as well as with MathWorks Simmechanics. However, some mechanical separation effects need very high sampling rates. Therefore this phenomenon needs a high-fidelity and a coarse model, one in the EFES and the other for the real-time closed loop. The Modelica version also contains the electrical command chain as triggered by Mission and Vehicle Management software (MVM). Because the modeling is physics based any conceivable hardware failure can easily be implemented at the very location where it occurs. A new, manual generation of differential equation is not necessary.

Fig. 5 illustrates the separation model.

As a second subsystem the attitude thrusters and the Vinci engine is modelled physically via EcosimPro (see fig 6). Contrary to SimMechanics, EcosimPro and Modelica have their own modeling software. Both are incorporated via the s-function mechanism of Matlab. This is not an ideal solution as it demands manual switching between three simulation environments. Because of the model heritage of EcosimPro for propulsion and the wide application of Modelica this drawback has been accepted.

Fig. 11 illustrates the requirement linking feature. Requirements can be written in DOORS or as Word file. A hyperlink can then point to the place of the Simulink software where the requirement is dealt with. The link is bi-directional. The advantage of this feature is that the programmer is always reminded of the link and will deal with potential software modification accordingly and carefully. The auto-report feature can extract the information and generates a well-formatted document. A second document is the 'Software Design Definition' which is built from the comments in the Simulink software (free comments, description of the connections, variables, etc.).

Not illustrated on the right side is a feature concerning testing and verification. In line with the MBD philosophy, which suggests that software designing, programming and testing should be performed as concurrently as possible, test points are placed at the appropriate places in the software right at the beginning. The test elements are triggered by violations of thresholds or perform statistical evaluations of signals etc. They can be selectively switched on or off.

The next section illustrates some application of the design suite.

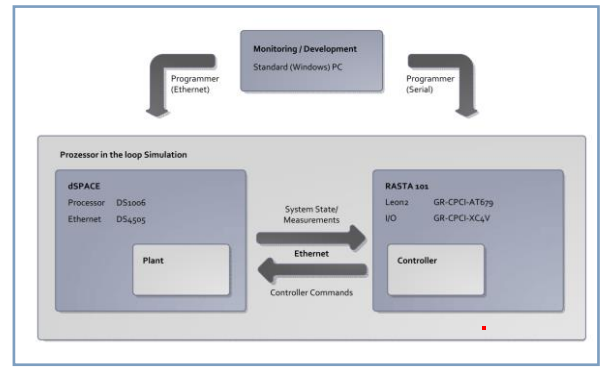

Fig 4. RASTA Configuration

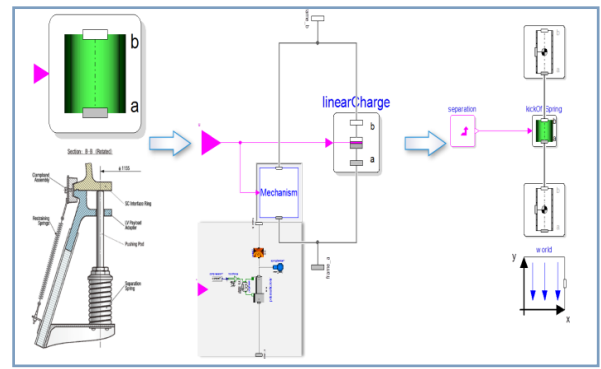

Fig 5. Separation Mechanism with Modelica

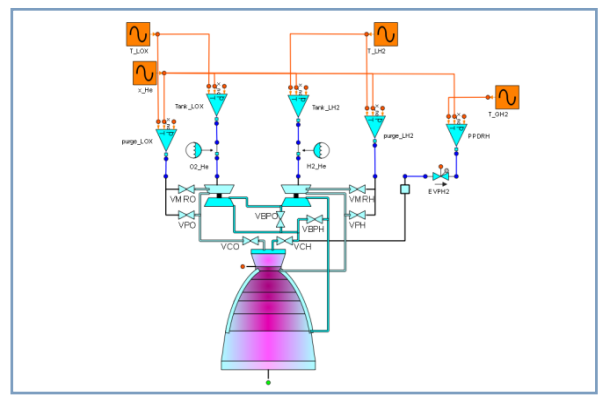

Fig 6. Vinci Engine with EcosimPro

## UPPER STAGE MISSION DESIGN WITH USACDF

The framework has been developed incrementally. Several benchmark design tasks have been studied in order to validate and improve the framework. In the following we will briefly present one case. Fig 6 illustrates an upper stage with one payload during a particular mission phase named 'long coasting phase'. Essentialy this is a barbecue mode of up to 5 hours with a spin rate of about 2 to 5 deg/sec. The special dynamics of nutation and precession can collect the propellant of a centrally placed tank in the form of a bulge (blue fluid in fig 7). Obviously a high dynamic unbalance (i.e. non-diagonal elements of the inertia matrix) will occur.

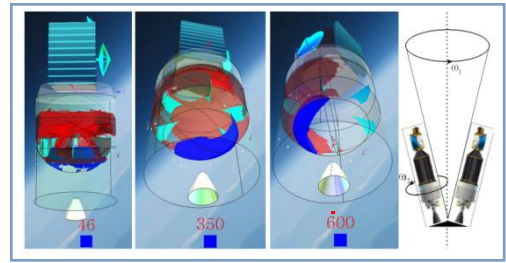
Fig 7. Spinning Stage with Sloshing Bulge

The long coasting phase with a high amount of propellant is a good example to illustrate the enhanced options, offered in an MBD framework, when it comes to advanced control algorithm design.

A model predictive approach (see [12], [13]) has been used to establish the optimal solution in terms of minimum number of actuation. A second adaptive solution has then been developed based on system identification of the propellant position and transformation of the control in the principal axis system. MBD allows exploring advanced concepts easily. Modern control concepts most often rely on the availability models including uncertainties. They are readily available in the MBD framework to various degree of details.

On-board re-configurability is illustrated in fig.8. The mission and vehicle management has been implemented with a state machine concept via MathWorks Stateflow toolbox.

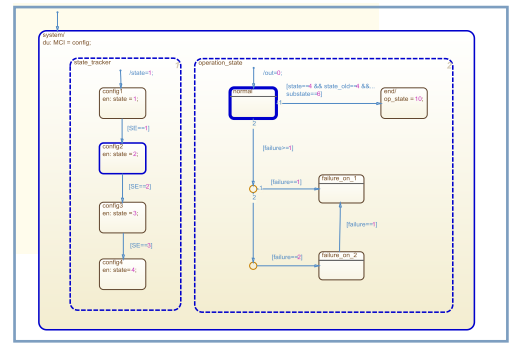
Fig 8. StateFlow Implementation of MVM

A state machine based expression of the mission sequence provides a very transparent way of defining branches at any place in the nominal sequence when a failure occurs. Emergency releases are triggered when an excessive propellant depletion is detected. Depending on when the failure occurs different actions have to be taken. The use of state machine syntax as opposed to a convolute of *if, then, else, case,* constructs is more reliable in terms of avoiding programming errors.

Fig 9 illustrates a complete payload release sequence expressed in angular rates. Two payloads are released. The supporting structure is released in between. The framework also contains Monte Carlo, Worst Case and Cross Entropy features (see [14]). Finally Fig 10 illustrates the tree like composition of the requirements .

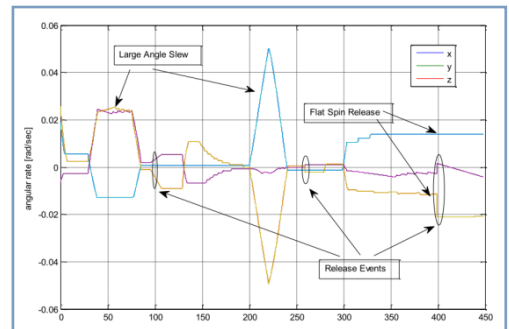
Fig 9. Angular Rate of a two Payload Release

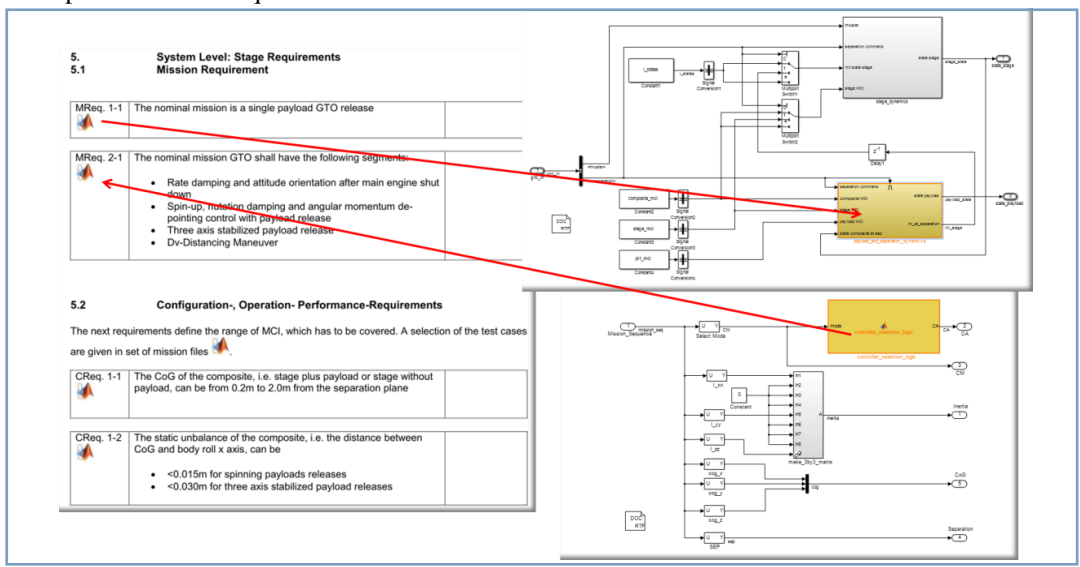
Fig 10. Tree-like Requirement Specification with two-way linking to Simulink Implementation

# CONCLUSION

The current framework has been designed to answer cross disciplinary system wide questions related to the safety and efficiency of complex space operations. Using coupled GNC CFD simulations and analyses permitted the prediction of nonlinear flight mechanics evolutions related to tank fluids motions in micro gravity flight. By closing the loop it was possible to assess nutation stability boundaries of the current system. Furthermore, the framework has allowed taking adequate design measures using model predictive control strategies for optimizing the actuation profile under stability and mission constraints such as roll reversals and spinned separations.

Integrating adequate physical propulsion models allows further to perform detailed GNC propulsion FDIR analyses for the development and sizing of adequate reconfiguration strategies.

Furthermore, efficient and safe payload separation strategies have been integrated into the MVM and GNC simulation tools. It is now possible to optimize upper stage mission sequences respecting timing, plume impingement as well as collision avoidance constraints. New worst case and stochastic verification strategies have been developed to better understand the limitations of existing designs as well as to be able to explore the limits of performance by adopting advance control concepts without too much conservatism.

The subsequent project follow-up will concentrate on the experimental capabilities of the USACDF framework. This will include experimental modeling, control and uncertainty validation techniques applied on various flight benchmarks and laboratory test facilities in order to assess innovative designs and to accelerate the TRL maturation process which is the only way to develop in time competitive technologies.

# ACKNOWLEDGMENT

The work was been performed within ESA's Future Launcher Preparatory Program (FLPP). Part of the works has been carried out by DLR Institute of Space Systems, Bremen, Germany and Astos Solustions GmbH, Unterkirnach, Germany. The authors thank Maxime Perrotin for the support concerning TASTE and all the reviewers of ESTEC TEC-ECN for sharing their thoughts and time.